\documentclass[epj,final]{svjour}

\usepackage{graphics}

\newcommand{\eg} {{\it e.g., }}
\newcommand{\eps} {\varepsilon}
\newcommand{\half} {\frac{1}{2}}
\newcommand{\ie} {{\it i.e., }}
\newcommand{\rmd} {{\rm d}}

\begin{document}

\title{Swelling kinetics of the onion phase}

\author{Haim Diamant\inst{1,2,3}\thanks{Corresponding author
at The University of Chicago, \email{diamant@control.uchicago.edu}}
\and Michael E. Cates\inst{2}}

\institute{School of Physics \& Astronomy, Raymond \& Beverly
Sackler Faculty of Exact Sciences, Tel Aviv University, 69978 Tel
Aviv, Israel \and Department of Physics \& Astronomy, The
University of Edinburgh, King's Buildings, Mayfield Road,
Edinburgh EH9 3JZ, Scotland \and The James Franck Institute, The
University of Chicago, 5640 South Ellis Avenue, Chicago, IL 60637,
USA}

\date{Received: date / Revised version: date}

\abstract{A theory is presented for the behavior of an array of
multi-lamellar vesicles (the onion phase) upon addition of
solvent. A unique feature of this system is the possibility to
sustain pressure gradients by tension in the lamellae. Tension
enables the onions to remain stable beyond the unbinding point of
a flat lamellar stack. The model accounts for various
concentration profiles and interfaces developing in the onion as
it swells. In particular, densely packed `onion cores' are shown
to appear, as observed in experiments. The formation of interfaces
and onion cores may represent an unusual example of stabilization
of curved interfaces in confined geometry.
}

\maketitle


\section{Introduction}
\label{sec_intro}

The self-assembly of amphiphilic molecules (surfactants)
in solution has been the subject of intensive research in the
past decades \cite{Benshaul,GompperSchick}.
Such molecules may self-assemble into a wide variety of aggregate
morphologies such as micelles, bilayers,
bilayer stacks (the lamellar, or L$_\alpha$ phase) and
various other liquid-crystalline structures.
The richness of these phenomena is both useful
for numerous chemical and biochemical applications and
challenging for basic research.

Yet another morphology of amphiphilic self-assembly has recently
drawn considerable attention.
When a lamellar (L$_\alpha$) phase is subjected
to shear, it undergoes a dynamic transition into an array
of close-packed multilayer vesicles, referred to as {\it
the onion phase}
\cite{Roux1,Roux2,Roux3,Roux4,Roux5}.
The typical size of an `onion' is of order 1--10 $\mu$m,
whereas the inter-membrane spacing is of order 10 nm, and the
thickness of the membrane itself is of order of a few nm.
Thus, each onion comprises a spherical stack
of more than several hundred spaciously packed membranes.
Although this structure is not an equilibrium one, it may
remain stable for days.
In addition, one can control the size of the onions by changing the
shear rate \cite{Roux1}, and encapsulate small particles inside them
\cite{encapsulation}. These appealing features have potential
applications, \eg in the pharmaceutical industry.

Various properties of the onion phase are not well understood
theoretically. In particular, the detailed mechanism of the
L$_\alpha$-to-onion dynamic transition has not yet been
established \cite{ZilmanGranek}.
Simple scaling arguments have been successful
in accounting for the dependence of the onion size on shear rate
and inter-membrane spacing \cite{Roux2,Lekker}, as well as the
viscoelastic behavior of the onion phase \cite{Panizza}.

In the current work we focus on another property of the onion
phase --- its behavior under dilution.
This aspect has been investigated in recent experiments
\cite{Buchanan1,Buchanan2}. When a regular L$_\alpha$
phase is diluted, the added solvent is accommodated in the
inter-membrane spacings. The membrane stack thereby swells, until
a `melting' transition into a disordered L$_3$ (`sponge') phase
occurs \cite{GompperSchick,Safran}.
The onion phase is observed to dissolve
into L$_3$ as well. Yet, because of the confined spherical
geometry, the stack cannot swell without a supply of additional
surfactant. (Amphiphilic membranes, to a good approximation, are
practically inextensible \cite{inextensible}.) As a
result, the dissolution progresses very slowly (over days)
through onion coalescence and disintegration, giving rise at
intermediate times to various defects and instabilities.
In particular, the formation of dense onion
cores has been observed in two different systems
\cite{Buchanan1,Buchanan2}, as is demonstrated in
Fig.~\ref{fig_core_exp}.

Our model of onion swelling relies on an assumption concerning
separation of time scales.
The time scale of swelling of the
entire onion is assumed to be much longer
than the one required for internal equilibration within the onion.
Experimentally, the former is of order of a few hours. The latter
is related to the time scale characterizing the formation of
passages (`necks') between adjacent membranes \cite{passages}.
We thus assume that passages form on time scales much
shorter than hours. Though reasonable, the validity of this
assumption is still to be established experimentally. There is a
third time scale, corresponding to onion coalescence and
disintegration into L$_3$.
Experimentally, it is found to be of order of days and will be
ignored here.
Thus, although the onion phase is a system far from
equilibrium, the above assumption allows us to present an essentially
equilibrium model of onion swelling.

A fundamental question arising from the study of onion dissolution
concerns the possibility to stabilize a curved interface between two
coexisting domains in confined geometry. In common situations such
a possibility does not exist. (For example, a liquid droplet
inside a coexisting vapor phase is unstable, and will either
shrink and vanish or expand to a macroscopic phase
\cite{Kittel}; this is the origin of nucleation barriers in
1st-order phase transitions.)
Stable interfaces in confined geometry are found in specific
systems, such as magnetic garnet films, block-copolymer melts,
phospholipid monolayers and microemulsions,
exhibiting modulated phases \cite{Andelman95}.
Competing interactions in such systems lead to a negative
surface tension (\ie negative stiffness term in a coarse-grained
Ginzburg-Landau free energy), which is stabilized by higher-order
terms. This gives rise to a finite characteristic length scale
of interface modulation \cite{Andelman95}.
As seen in Fig.~\ref{fig_core_exp}, onions under dilution seem
to exhibit a stable interface between a confined, dense core and
a dilute shell. We shall try to demonstrate
that this behavior might represent a new way to stabilize a
curved interface, arising from the unique ability of onions
to sustain pressure gradients at equilibrium.

In section~\ref{sec_tension} we extend a simple theory for the
unbinding transition in a flat L$_\alpha$ phase \cite{Milner92} to
the case of membranes with tension. Based on this extension, we
formulate in section~\ref{sec_onion} a model of a single onion as an
inhomogeneous, spherical membrane stack. We then present the 
resulting concentration and tension profiles in the onion. 
Finally, in section~\ref{sec_summary}, we summarize the
results and point at future directions.

\section{Lamellar Phase with Tension}
\label{sec_tension}

Fluid membranes in a lamellar
stack experience steric repulsion
arising from their reduced undulation entropy
(the Helfrich interaction) \cite{Helfrich1}.
In the case of tensionless membranes the interaction energy per
unit area is
\begin{equation}
  f_{\rm und}(D) = \frac{b T^2} {\kappa D^2},
\label{Helf_potential}
\end{equation}
where $T$ is the temperature (in energy units, \ie
$k_{\rm B}\equiv 1$), $\kappa$ the bending rigidity of the
membranes, and $D$ the inter-membrane spacing.
(The numerical prefactor $b$ is still under controversy
\cite{Sornette}; Helfrich's calculation
\cite{Helfrich1} gives $b=3\pi^2/128\simeq 0.2$,
whereas computer simulations \cite{Lipowsky89} yield a lower value
of $b\simeq 0.06$.)
The long range of the Helfrich interaction, 
Eq.~(\ref{Helf_potential}),
results from the `floppiness', \ie strong thermal undulations,
of tensionless membranes.
The interplay between the Helfrich repulsion and other, direct
interactions determines when a lamellar stack of membranes becomes
unstable and unbinds.
The rich critical behavior exhibited by this system was thoroughly
investigated using functional renormalization group
techniques \cite{Nelson_Leibler,LipowskyLeibler}.

Subsequently, a much simpler theory for the unbinding of a lamellar
membrane stack was proposed \cite{Milner92}.
It employs a similar argument to Flory's for polymers ---
since the stack is a `soft', entropy-dominated system, one has
to accurately account for entropy [\ie the Helfrich repulsion,
Eq.~(\ref{Helf_potential})], while the other interactions can be
incorporated in an approximate, 2nd-virial term.
The resulting (grand-canonical) free energy per unit
volume of the lamellar stack is
\begin{equation}
  f(\phi) = \half\phi^3 - \chi\phi^2 - \mu\phi,
\label{fflat}
\end{equation}
where $\phi\equiv\delta/D$ is the surfactant volume fraction
($\delta$ being the membrane thickness),
$\chi$ is a 2nd-virial coefficient, and $\mu$ the
surfactant chemical potential. All energy densities have been
scaled by $2bT^2/(\kappa\delta^3)$.
For $\chi>0$ the free energy (\ref{fflat}) describes a 1st-order
unbinding transition as $\mu$ is lowered.
The chemical potentials and volume fractions
corresponding to the binodal and spinodal of this transition are
\begin{eqnarray}
  \mu_{\rm bin}&=&-\chi^2/2,\ \ \ \
  \phi_{\rm bin}=\chi
\nonumber \\
  \mu_{\rm sp}&=&-2\chi^2/3,\ \ \
  \phi_{\rm sp}=2\chi/3.
\label{binspin_flat}
\end{eqnarray}
The free energy (\ref{fflat}) also has a critical point at
$\mu=\chi=0$, which is of much theoretical interest
\cite{Nelson_Leibler}, but of no relevance to the current
discussion; hereafter, a positive value of $\chi$ is assumed.

Consider now a stack of membranes having tension $\sigma$.
Surface tension strongly suppresses membrane undulations and,
hence, the fluctuation-induced interaction between tense membranes
has a much shorter range.
The calculation of this interaction is more complicated than
for tensionless membranes \cite{Sornette}.
Renormalization-group calculations \cite{RG_tension} and computer
simulations \cite{Netz95} yield an exponential
decay with distance.
A simpler, self-consistent calculation of this interaction
\cite{Seifert} gives
\begin{equation}
  f_{\rm und}(D) = \frac{bT^2}{\kappa D^2} \left[\frac{D/l_T}
  {\sinh(D/l_T)}\right]^2,
\label{potential_tension}
\end{equation}
where $l_T$ is the length arising from the combination of tension
and thermal energy, $l_T\equiv (2T/\pi\sigma)^{1/2}$. The
dimensionless parameter $x\equiv D/l_T=\delta/(l_T\phi)$,
depending on both $\sigma$ and $\phi$,
determines whether the tension has a significant effect on the
interaction. For $x\ll 1$ $f_{\rm und}$ coincides with the
tensionless expression, Eq.~(\ref{Helf_potential}), whereas for
$x\gg 1$ the tension strongly suppresses membrane undulations and
the interaction decays exponentially with
distance, in accord with the renormalization-group result
\cite{RG_tension}. (The numerical prefactor in $l_T$ was chosen so
as to recover the renormalization-group result for high tension.)
For brevity we use hereafter
the following notation:
\begin{eqnarray*}
  G(x) &\equiv& -\frac{1}{2\sinh^2 x},\ \ \
  g(x) \equiv \frac{\rmd G}{\rmd x} = \frac{\cosh x}{\sinh^3 x},
  \nonumber\\
  g'(x) &\equiv& \frac{\rmd g}{\rmd x} = \frac{4\sinh^2 x - 3}
  {\sinh^4 x}.
\end{eqnarray*}

The `Flory-like' free energy per unit volume in the tense case is
\begin{equation}
  f(\phi,\sigma) = -x^2 G(x) \phi^3 - \chi\phi^2 - \mu\phi,
\label{ftense}
\end{equation}
where the energy densities have been scaled, again, by
$2bT^2/(\kappa\delta^3)$. (Note that $x$ should not be taken as an
independent degree of freedom but as $\phi$-dependent. The two
independent degrees of freedom are $\phi$ and $\sigma$.
Nevertheless, $x$ is extensively used below, in order to make the
formulation more concise.)

Due to the additional degree of freedom --- membrane tension ---
bound stacks can be stabilized even beyond the unbinding point of
the tensionless case, \ie for $\mu<\mu_{\rm sp}$. Hence, instead
of a transition point there is a transition line, $\sigma^*(\mu)$,
such that the stack is bound for $\sigma>\sigma^*$, and unbound
for $\sigma<\sigma^*$. The equations for the binodal and spinodal
lines are:
\begin{eqnarray}
  &\mbox{binodal:} \ \ \ &
  2\frac{G(x)+xg(x)}{[x^2g(x)]^2} = \frac{\mu}{\mu_{\rm bin}},
  \ \ \ \ \phi = \frac{\chi}{x^3g(x)}
\nonumber\\
  &\mbox{spinodal:} &
  6 \frac {G(x)-xg(x)-x^2g'(x)}{[x^3g'(x)]^2} =
  \frac{\mu}{\mu_{\rm sp}},
\nonumber\\
  && \phi = -\frac{2\chi}{x^4g'(x)}.
\label{tense_transition}
\end{eqnarray}
The two lines are drawn in Fig.~\ref{fig_tense_diag}A. Figure
\ref{fig_tense_diag}B shows the line $x^*(\mu)$ corresponding to
the binodal. When $\mu$ is slightly lower than $\mu_{\rm bin}$,
the value of $x^*$ required to stabilize the stack increases
sharply. On the other hand, values of $x^*$ much larger than 1 are
required only for chemical potentials much lower than $\mu_{\rm
bin}$. Hence, realistic values for $x^*$ should be of order
0.1--3.

\section{Onion Swelling}
\label{sec_onion}

\subsection{The Model}
\label{sec_model}

In view of the assumption regarding separation of time scales,
presented in section~\ref{sec_intro}, we consider the onion as a
spherical stack of {\it constrained size}. (Recall that the
initial size of an onion is, by itself, determined not by
thermodynamic equilibrium but by the shear rate that led to its
formation \cite{Roux1}.) As dilution progresses, the constrained
radius, $R$, increases slowly, such that at any instant the
interior of the onion can be assumed in thermodynamic equilibrium.
This implies, in particular, that the entire onion has a single,
uniform chemical potential, $\mu$. Since the temperature, $T$, and
the total number of surfactant molecules in the onion, $N$, are
taken as fixed, $\mu$ must continuously decrease upon swelling.
Thus, the equilibrium ensemble relevant to the actual dilution
process (at time scales shorter than hours) is that of fixed
$(T,R,N)$, where $R$ is regarded as a slowly increasing 
external parameter, leading to a slow decrease in
$\mu$ and $p_0$, the osmotic pressure. For mathematical
convenience, the model is formulated in the equivalent ensemble of
fixed $(T,p_0,\mu)$, where $\mu$ is is a slowly decreasing
parameter.

As mentioned in section~\ref{sec_intro}, each onion comprises a
stack of many spaciously packed membranes. It is therefore
justified to employ a coarse-grained, continuum model. Unlike a
flat lamellar phase, the inter-membrane spacing in the onion is
not expected to be uniform. Hence, we allow for non-uniform
profiles, writing a Ginzburg-Landau free energy of the form
\begin{equation}
  F[\phi(\vec{r}),\sigma(\vec{r})] =
  \int\rmd\vec{r} \left[ \half\Omega |\nabla\phi|^2
  + f(\phi,\sigma) \right] + \frac{4\pi}{3} p_0 R^3.
\label{F_GL}
\end{equation}
The free energy density, $f(\phi,\sigma)$, has been defined in
equation~(\ref{ftense}), and the
integration is over the volume of the onion. An energetic penalty
for concentration gradients has been included in equation~(\ref{F_GL}),
where $\Omega$ is a stiffness coefficient. The Lagrange multiplier
$p_0$ should ensure that the onion radius has the constrained value
$R$.

The state of the onion is defined by the concentration profile
$\phi(r)$ (or, equivalently, by the set of inter-membrane
spacings). The equilibrium concentration profile is thereby found
from a variation principle.
A major theoretical complication is the fact that the tension in the 
membranes enters at two distinct levels---as a macroscopic
parameter, \eg balancing the pressure difference across a membrane,
and as a microscopic parameter having a dramatic effect on
membrane interaction.
In order to overcome this obstacle we use a self-consistent
scheme.
We assume that the membranes have attained certain values of tension,
leading to a (still unknown) profile $x(r)$. Subsequently, we find
the resulting concentration and pressure profiles and then require
self-consistency, \ie that the presumed tension in the membranes
balance the pressure gradients. Note that this self-consistency
does not require that the tension profile be smooth. Hence, no
penalty for spatial changes in tension has been included in
equation~(\ref{F_GL}); near-by membranes may have very different
tensions. In practice, the formation of passages should act to
equalize the tension between layers and, hence, a certain penalty
for tension differences is expected. We assume that such a
tension-gradient term would not have a drastic effect on the
results. Hence, we omit it to avoid addition of another parameter
to the model.

\subsection{Profile Equations}
\label{sec_profeq}

By taking the variation of $F$ with respect to $\phi(r<R)$ we
obtain the first profile equation,
\begin{equation}
  \Omega\nabla^2\phi + x^2[G(x)-xg(x)]\phi^2 + 2\chi\phi
  + \mu = 0,
\label{profile1}
\end{equation}
where, for our spherical-symmetric case, $\nabla^2=\rmd^2/\rmd
r^2+(2/r)\rmd/\rmd r$. (Recall that the variation is taken while
keeping $\sigma$, not $x$, fixed.)

Equation~(\ref{profile1}), obtained from a variation of a
Ginzburg-Landau functional, is of the generic form widely used
to study interfaces \cite{Langer}.
It is equivalent to imposing a uniform chemical potential
throughout the system.
Such a profile equation is usually supplemented by
boundary conditions for the order parameter and its gradient
far away from the interface, that ensure the
uniformity of pressure.
Such a system of equations is over-determined (a 2nd-order
equation with four boundary conditions). It has a solution
for a flat geometry, but does not have one for a confined
(\eg spherical) geometry. Thus, such a Ginzburg-Landau
formalism cannot in general produce stable, confined
interfaces.
However, the onion phase has a very special
property. Since the system is composed of concentric closed
sheets, {\em the pressure need not be uniform throughout the
system}; pressure gradients can be sustained by the tension in the
membranes. In particular, if the system is divided into coexisting
domains, the pressure does not have to be equal
in the different domains. This property bypasses the
usual uniform-pressure boundary conditions.
They are to be replaced by an equation, derived below,
balancing pressure gradients and tension.

Using Green's identity
and equation~(\ref{profile1}), we can rewrite the free energy at
equilibrium as
\begin{eqnarray}
  F &=& -\int\rmd\vec{r} \half\phi
  \{ x^2[G(x)+xg(x)]\phi^2 + \mu \}
\nonumber\\
  & & + 4\pi R^2 \half \Omega \phi(R)\frac{\rmd\phi(R)}{\rmd r}
  + \frac{4\pi}{3} R^3 p_0,
\label{Feq}
\end{eqnarray}
and identify the integrand as the local pressure,
\begin{equation}
  p = \half\phi \{ x^2[G(x)+xg(x)]\phi^2 + \mu \}.
\label{pressure}
\end{equation}
Local balance between the pressure gradient and membrane tension
is accounted for by a Laplace equation,
\begin{equation}
  \hat{\sigma}_i = \half r_i (p_{i-1}-p_{i})
  \simeq -\half r D(r) \frac{\rmd p}{\rmd r},
\label{selfconsistency}
\end{equation}
where $\hat{\sigma}_i$ is the rescaled tension (having dimension of
length) of membrane $i$, $r_i$ its radius, and $p_i$ the pressure
just outside it.
In the second equality we have assumed that the pressure profile is
a smooth function on the length scale of $D$, the inter-membrane 
spacing.
Equation~(\ref{selfconsistency}) ensures that the pressure
gradients resulting from the concentration profile are consistent
with the presumed tension profile.
For the sake of mathematical convenience, we shall treat the tension
profile as a smooth function as well, \ie represent spatial variations
of $\sigma$ by a first derivative, $\rmd\sigma/\rmd r$.
Since the model allows for sharp changes in $\sigma$, this
approximation is merely technical and not physically corroborated.
Consequently, the sharp changes will show up as singularities
in an otherwise smooth tension profile, and will have to be treated
separately to ensure that the smoothness of $\phi$ and $p$ is
maintained.
When equation~(\ref{pressure}) is substituted in equation~(\ref{selfconsistency}),
the smoothness assumption leads to the following, second profile
equation:
\begin{eqnarray}
  && [2G(x)+4xg(x)+x^2g'(x)]\frac{1}{x}
  \frac{\rmd x}{\rmd r}
\nonumber\\
  && + \{3[G(x)+xg(x)]+\frac{\mu}{x^2\phi^2}\}
  \frac{1}{\phi} \frac{\rmd\phi}{\rmd r} +
  \frac{4\eps}{r} = 0,
\label{profile2}
\end{eqnarray}
where $\eps\equiv\kappa/(\pi bT)$.

The profile equations (\ref{profile1}) and (\ref{profile2}), along
with appropriate boundary conditions, determine the profiles
$\phi(r)$ and $x(r)$.

\subsection{Boundary Conditions}
\label{sec_boundary}

The onion is in contact with the surrounding environment through
its outer layer. This layer requires a separate treatment, so as
to yield the boundary conditions for the profile equations
(\ref{profile1}) and (\ref{profile2}), \ie $\phi(R)$, $x(R)$ and
$\rmd\phi(R)/\rmd r$.
We employ a self-consistent scheme similar to the one of 
section \ref{sec_profeq},
\ie using a variation principle for $\phi(R)$ and requiring
that $\sigma(R)$ balance the pressure difference across the 
outer membrane.

Discretization of the integral in equation~(\ref{F_GL}) and
taking the variation of $F$ with respect to $\phi(R)$
give the boundary condition for the concentration gradient,
\begin{equation}
  \frac{\rmd\phi(R)}{\rmd r} = -\left.\frac{\delta}{\Omega}
  \{x^2[G(x)-xg(x)]\phi^2 + 2\chi\phi + \mu\}\right|_{r=R}.
\label{bc1}
\end{equation}
Variation of $F$ [Eq.~(\ref{Feq})] with respect to $R$ yields the
mechanical equilibrium (Laplace) equation for the outer membrane,
\begin{equation}
  \half \Omega \phi(R) \frac{\rmd\phi(R)}{\rmd r} =
  \half R \{p[\phi(R),\sigma(R)] - p_0\},
\label{outer_Laplace}
\end{equation}
where $p[\phi(R),\sigma(R)]$ is the pressure just inside the outer
sheet, whose dependence on $\phi(R)$ and $\sigma(R)$ is given by
equation~(\ref{pressure}). 
The multiplier $p_0$, coupled to the total volume 
[see equation (\ref{F_GL})], is the external osmotic pressure 
exerted on the onion, \ie the pressure just outside the outer 
sheet.
Hence, the left-hand side of equation~(\ref{outer_Laplace}) is 
identified as the tension of the outer layer, $\hat{\sigma}(R)$. 
This observation, together with equation~(\ref{bc1}) and the 
definition of $x$, $\hat{\sigma}=\eps\delta x^2\phi^2$, lead
to two equations for $x(R)$ and $\phi(R)$,
\begin{eqnarray}
  && x^2[G(x)-xg(x)]\phi^2 + 2(\chi+\eps x^2)\phi
  + \mu = 0
\label{bc02} \\
  && x^2[G(x)+xg(x)]\phi^2 - \frac{4\eps\delta}{R}
  x^2\phi + \mu - \frac{2p_0}{\phi} = 0.
\label{bc03}
\end{eqnarray}

One can proceed using boundary conditions (\ref{bc02}) and
(\ref{bc03}). Yet, the formulation can be further simplified by
employing an additional assumption. Prior to dilution, the onion
is in contact with the outer membranes of neighboring onions in
the close-packed phase.
This
situation persists as long as the unbinding point of the regular
L$_\alpha$ phase has not been reached. During this initial stage
of dilution the inter-membrane spacing inside the onion is equal
to the spacing between the outer membranes of neighboring onions
(\ie $\phi$ is uniform throughout the sample). Consequently, the
pressure is equal on both sides of the outer membrane and its tension
vanishes, $x(R)=0$. The profile equations (\ref{profile1}) and
(\ref{profile2}) then have the trivial uniform solution $x(r)\equiv 0,
\phi(r)\equiv\phi(R)$. When the unbinding point is reached,
$\mu<\mu_{\rm bin}$, the onion becomes separated from the
surrounding membranes. The undulation pressure is still exerted on
the outer membrane from inside but vanishes outside, and tension
must appear. Since the external pressure should become very low
compared
to the internal one, we may neglect $p_0$ and assume that the
internal pressure is balanced primarily by tension. This
assumption leads to the following simplified boundary conditions:
\begin{eqnarray}
  && \phi(R) = \left. \frac{\chi(1+\rho x^2)}{x^3g(x)}
  \right|_{r=R}
\label{bc2}
\\
  && \left. \frac{ 2[G(x)+xg(x)] (1+ \rho x^2)^2 } {[x^2g(x)]^2}
  - \frac{8\eps\delta}{\chi R} \frac{1+\rho x^2}{xg(x)}
  \right|_{r=R} \nonumber\\
  && = \frac{\mu}{\mu_{\rm bin}},
\label{bc3}
\end{eqnarray}
where $\rho\equiv(\eps/\chi)(1+2\delta/R)$.

Comparison between the boundary conditions
(\ref{bc2})--(\ref{bc3}) and equation~(\ref{tense_transition}) shows
that for $\mu\rightarrow\mu_{\rm bin}^-$ the boundary values,
$x(R)$ and $\phi(R)$, coincide with the binodal ones, \ie
$x(R)\rightarrow 0$, $\phi(R)\rightarrow\phi_{\rm bin}$, and
$\rmd\phi(R)/\rmd r\rightarrow 0$. Thus, the outer layer {\it
continuously} acquires different features as the membranes
surrounding the onion unbind. The continuous departure of $x(R)$
from zero is also demonstrated in Fig.~\ref{fig_xR}. The increase
of $x(R)$ when $\mu$ becomes slightly lower than $\mu_{\rm bin}$
is found from equation~(\ref{bc3}) to scale like
\begin{equation}
  x(R) \sim \left|\mu-\mu_{\rm bin}\right|^{1/2},
  \ \ \ \ \mu\rightarrow\mu_{\rm bin}^-.
\end{equation}
Hence, the tension in the outer layer, $\sigma(R)\sim x^2(R)$,
increases linearly with $|\mu-\mu_{\rm bin}|$.
In addition, as $\mu$ is further decreased, it is verified
that $x(R)$ of equation~(\ref{bc3}) always remains
smaller than the binodal value of equation~(\ref{tense_transition}),
required to stabilize bound membranes.
The outer membranes of different onions, therefore,
remain unbound from one another throughout the dilution.

\subsection{Profiles and Interfaces}
\label{sec_profile}

In order to calculate concentration and tension profiles in the
onion, one should solve the profile equations (\ref{profile1}) and
(\ref{profile2}) subject to the boundary conditions (\ref{bc1}),
(\ref{bc2}) and (\ref{bc3}). This is a system of coupled
nonlinear equations, which is solved numerically.
Note, however, that the entire formulation given
above applies only in the case of low surfactant concentration
($\phi\ll 1$). In order to correctly account for concentrated
domains in the onion ($\phi\simeq 1$) we actually solved a
modified, more complicated set of equations, as presented in the
Appendix. Before considering the
numerical results, it is useful to examine some general features
of the solutions.

Figure \ref{fig_pressure} shows the $x$ dependence of the local
pressure for a given $\phi$ [cf.\ Eq.~(\ref{pressure})].
There are two stationary points of the pressure as a function of
$x$: $x=0$ and $x=x_1\simeq 1.17$. They correspond to two
singularities of equation~(\ref{profile2}) for the tension profile.
Recall that the model allows for sharp changes in $x(r)$, as long
as the pressure profile remains smooth. The singularities in the
tension profile signal such jumps as a consequence of the
smoothing assumption in equation~(\ref{profile2}); they must be treated
separately, as follows.
If the profile reaches
the singularity $x(r)=x_1$, it cannot jump to any other value of
$x$ without violating the smoothness of $p$. Since
equation~(\ref{profile2}) is 1st-order, this implies that the rest of
the profile must stay at $x_1$. A different behavior is expected
if the singular point $x(r)=0$ is reached. In this case the
profile cannot remain, in general, at $x=0$ (zero tension), since
this would imply also uniform pressure and, hence, a uniform
concentration profile. The available option for the next membrane
is to jump to a value of $x=x_2\simeq 1.91$ while maintaining a
continuous pressure profile. The meaning of the latter observation
is that {\it a sharp interface in the tension profile, between a
low-tension domain ($x\sim 0$) and a high-tension one ($x\sim 1$),
is permitted}.

We now turn to the results of the numerical analysis.
The analysis is restricted to stages of dilution where
$\mu$ has not reached very low values, so that $x(R)$ could be
assumed smaller than $x_1$ (see, \eg the values of $x(R)$ in
Fig.~\ref{fig_xR}). The solutions
to the profile equations are divided in this case into three
families:
\begin{itemize}

\item[(i)] profiles which descend from $x(R)$ to $x=0$, then jump
sharply to $x=x_2$ and descend to remain at $x=x_1$
(Fig.~\ref{fig_belt})

\item[(ii)] profiles which, like family (i), have a sharp jump from
$x=0$ to $x=x_2$, but then proceed to higher values of $x$
(Fig.~\ref{fig_core})

\item[(iii)] smooth profiles which ascend from $x(R)$ to
$x=x_1$ and remain there (Fig.~\ref{fig_uniform}).

\end{itemize}

Family (iii) may be called `uniform', as it does not exhibit a
sharp jump in tension. Indeed, since the pressure monotonously
increases when going into the onion and $x$ is uniform throughout
the inner part of the onion, the concentration must increase as
well and not remain uniform. This increase, however, is only
logarithmic in $r$ [as can be found from equation~(\ref{profile2}) with
constant $x$]. The `belt' family (i) is mostly uniform as well ---
it has only a narrow region of relatively high tension. Most
interesting is the `core' family (ii). In this case the jump in
$x$ divides the onion into an outer, low-tension region, and an
inner, tense one. As we go inward, the tension continues to
increase until diverging at a finite radius, where the membrane
stack reaches the maximum concentration of close packing
($\phi=1$).
(Note that the close-packing limit $\phi\rightarrow 1$ does not
imply a divergent free energy, since it is accompanied by
$\sigma\rightarrow\infty$ (see Fig.~\ref{fig_core}).
The high tension exponentially suppresses the appropriate
term in the free energy [cf.\ Eq.~(\ref{f_tense_app})].)

The parameter space for numerical study is vast. After rescaling
all distances with $R$, we are left with five parameters: $\chi$,
$\eps$, $\delta$, $\Omega$, and $\mu$. Nonetheless, the
qualitative features, such as the three families of solutions
described above, are found to be robust over a wide range of
parameter values. Figure \ref{fig_profile_diagram} shows examples
of `profile diagrams' for three cuts through the parameter space.
Dilution beyond $\mu_{\rm bin}$ usually results first in belt
profiles. Cores may subsequently appear and, finally, the profile
shifts to the uniform family for low enough $\mu$. The structure
of the diagram, however, may be more complicated than this simple
sequence (Fig.~\ref{fig_profile_diagram}B). Note the wide range of
dilution over which onion cores may be stable. Since $\mu$ is not
expected in practice to become much lower than a few times
$\mu_{\rm bin}$, cores may remain stable over the entire process
of dissolution (as indeed observed in experiments \cite{Buchanan1}).
Not surprisingly, we find that core formation is promoted
(\ie occurs at higher $\mu$) by lower membrane rigidity or higher
temperature (smaller $\eps\sim\kappa/T$), and stronger
attraction between membranes (larger $\chi$).

The values assigned to $\Omega$ in Fig.~\ref{fig_profile_diagram}
are rather large (a particularly large value was taken in
Fig~\ref{fig_profile_diagram}B in order to demonstrate a richer
diagram). Diagrams for smaller $\Omega$ still exhibit the
belt-to-core transition, yet spatial variations occur on shorter
distances from the boundary, resulting in bigger `cores'. Another
source of quantitative uncertainty is the factor $b$ entering the
definition of $\eps$ [$\eps\equiv(\kappa/T)/(\pi b)$]. With $b$ in
the range between 0.06 (simulation \cite{Lipowsky89}) and 0.2
(theory \cite{Helfrich1}), one needs $\kappa/T<\sim 0.1$--0.5 in
order to get $\eps<\sim 0.5$, as required for core formation for
reasonable values of $\mu$ (Fig.~\ref{fig_profile_diagram}). These
values are lower than the ones expected in the relevant
experimental systems ($\kappa\simeq$ a few $T$). It should be
stressed, however, that we seek in this work merely qualitative
mechanisms, rather than an accurate predictive capability.

The narrow shell of high tension appearing in both the `belt' and
`core' profile families does not have, within the current
approximation, a significant effect on the
concentration profile (see Fig.~\ref{fig_belt}). Hence, unless an
experiment is devised which will be sensitive to membrane tension,
rather than density, this feature might not be directly
observable in experiments. Nevertheless, the highly tense
membranes may affect the {\em kinetics} of dissolution (\eg hinder
or assist water penetration and passage formation).
This might explain the
experimental observation of concentric `cracks' or `rings' in
diluted onions \cite{Buchanan2}.

\section{Summary}
\label{sec_summary}

We have presented a theory for the swelling of the onion phase
upon addition of solvent. A membrane stack in a spherical onion
configuration can remain stable far beyond the point where a flat
lamellar phase disintegrates. This stability is achieved due to a
tension profile acquired by the stack. As a result, the eventual
dissolution of individual onions is bound to rely on membrane
breakage and coalescence, thus taking very long time.

At the unbinding point of the regular L$_\alpha$ phase, when
individual onions become separated, a concentric shell of
membranes with high tension (`belt') should first appear in the onion.
If the membranes are not too rigid, subsequent formation of a dense
core might occur, as was observed in experiments. The cores may remain
stable under extensive dilution (cf.\ Fig.~\ref{fig_profile_diagram}).
In other cases, they may
eventually disappear, giving rise to more uniform profiles.

The interface formed between the inner, tense part of the onion,
beyond the `belt', and its outer part is {\em thermodynamically
stable} (on time scales shorter than hours) --- the chemical
potentials in the two domains are equal, and the pressure
gradient is exactly balanced by an appropriate tension profile.
This is a special example where an interface can be stabilized
in a confined
geometry without resorting to competing interactions.
Stability becomes possible due to the tension in the membranes,
which allows the system to avoid the usual equal-pressure
condition of coexistence. In the absence of a length scale arising
from competing interactions, the size of the inner domain must
scale with the onion radius. The proportionality factor, however,
may be small (see Fig.~\ref{fig_core}).

The conclusions drawn from the model rely on numerical solution of
the profile equations and boundary conditions for specific values
of parameters. Nevertheless, the general behavior presented in
section~\ref{sec_profile} (\eg the shift between the three families
of profiles) is found to be robust under change of parameters and
even certain changes of the boundary conditions. Hence, we believe
that the qualitative mechanisms indicated by this model are fairly
general. The system of coupled nonlinear profile equations derived
in section~\ref{sec_profeq} may, in principle, produce a much wider
variety of solutions. It should be interesting, therefore, to
explore further (possibly less physical) areas of the parameter
space.

The theory presented here should be regarded as a
preliminary step towards understanding the dissolution of the
onion phase. In particular, it is focused on the first stages of
dilution, where individual onions swell while maintaining
their integrity.
Further stages of dissolution involve onion breakage and coalescence,
where intriguing instabilities are observed \cite{Buchanan2}.
These phenomena probably require an altogether different theoretical
approach.

\begin{acknowledgement}
We benefited from discussions with
D.\ Andelman, M.\ Buchanan, J.\ Leng and T.\ A.\ Witten.
HD would like to thank the British Council and Israel
Ministry of Science for financial support, and the
University of Edinburgh for its warm hospitality.
\end{acknowledgement}

\section*{Appendix: Expressions for High Concentration}
\renewcommand{\theequation}{A\arabic{equation}}
\setcounter{equation}{0}

The various expressions derived in the previous sections apply
only in the limit of low surfactant volume fraction, \ie when the
inter-membrane spacing is much larger than the membrane thickness,
$\phi=\delta/D\ll 1$. Onion cores, however, are regions of
close-packed membranes, $\phi\simeq 1$. Hence, a reliable
calculation of profiles and profile diagrams, such as those
presented in Figs.\ \ref{fig_belt}--\ref{fig_profile_diagram},
requires modified equations, which are valid for high volume
fractions as well. The following Appendix presents these modified
expressions.

The Helfrich interaction between tensionless membranes
\cite{Helfrich1}, Eq.~(\ref{Helf_potential}), is readily
generalized to the case of finite membrane thickness.
One uses the same arguments, yet the
available space for undulations is now $D-\delta$ instead of $D$.
The resulting interaction energy per unit area is
\begin{equation}
  f_{\rm und}(D) = \frac{bT^2}{\kappa(D-\delta)^2}.
\end{equation}
This leads to the following `Flory-like' free energy density for a
tensionless stack,
\begin{equation}
  f(\phi)=\frac{\phi^3}{2(1-\phi)^2} - \chi\phi^2 -\mu\phi,
\end{equation}
which replaces equation~(\ref{fflat}). The
modified expressions for the binodal and spinodal arising from
this free energy are [compare to equation~(\ref{binspin_flat})]
\begin{eqnarray}
  \mu_{\rm bin} &=& -(\chi/2)\phi_{\rm bin}(1+\phi_{\rm bin}),
  \ \ \phi_{\rm bin}(1-\phi_{\rm bin})^{-3}=\chi
\nonumber \\
  \mu_{\rm sp} &=& -\chi\phi_{\rm sp} (1+4\phi_{\rm sp}/3 -
  \phi_{\rm sp}^2/3), \nonumber\\
  & & \phi_{\rm sp}(1-\phi_{\rm sp})^{-4}=2\chi/3.
\end{eqnarray}

Seifert's calculation of the fluctuation-induced interaction in
the presence of tension \cite{Seifert} is readily extended as
well. The only modification required in
equation~(\ref{potential_tension}) is the replacement of $D$ with
$D-\delta$. The resulting `Flory-like' free energy for a tense
stack [generalizing Eq.~(\ref{ftense})] is
\begin{equation}
  f(\phi,\sigma) = -x^2 G(x) \frac{\phi^3}{(1-\phi)^2} 
  - \chi\phi^2 - \mu\phi,
\label{f_tense_app}
\end{equation}
with a modified definition of $x$,
$x\equiv(D-\delta)/l_T=(\delta/l_T)(1-\phi)/\phi$.

Writing a Ginzburg-Landau free energy similar to equation~(\ref{F_GL})
and taking the variation with respect to $\phi(r<R)$, we obtain
the modified version of the first profile equation [which replaces
equation~(\ref{profile1})],
\begin{equation}
  \Omega\nabla^2\phi + x^2[G(x)-xg(x)/(1-\phi)]
  [\phi/(1-\phi)]^2 + 2\chi\phi   + \mu = 0.
\end{equation}
Rewriting the free energy after minimization [cf.\ Eq.
(\ref{Feq})], we identify the local pressure as
\begin{equation}
  p = \half\phi \{ x^2[G(x)+xg(x)/(1-\phi)][\phi/(1-\phi)]^2 + \mu \},
\label{pressure2}
\end{equation}
which replaces equation~(\ref{pressure}).\footnote
{Note that the stationary point of $p$ as function of $x$ for
fixed $\phi$, defined in section \ref{sec_onion} as $x=x_1$ 
(cf.\ Fig.~\ref{fig_pressure}), is no longer a constant but depends 
on the value of $\phi$. The same is true for $x=x_2$, for which
$p(x=x_2)=p(x=0)$.}
Substituting the modified expression for the local pressure,
Eq.~(\ref{pressure2}), in the self-consist\-ency condition,
Eq.~(\ref{selfconsistency}), we get the modified profile equation
[compare to equation~(\ref{profile2})],
\begin{eqnarray}
  && \frac{2(1-\phi)G(x)+(4-\phi)xg(x)+x^2g'(x)} {x(1-\phi)}
  \frac{\rmd x}{\rmd r} \nonumber\\
  && + \left\{ \frac{[2+(1-\phi)^3](1-\phi)G(x)+[2+(1-\phi)^2]xg(x)}
  {(1-\phi)^4} \right. \nonumber\\
  && + \left. \frac{(1-\phi)^2\mu}{x^2\phi^2} \right\}
  \frac{1}{\phi} \frac{\rmd\phi}{\rmd r}
  + \frac{4\eps}{r} = 0.
\end{eqnarray}

Finally, repeating the calculations described in
section~\ref{sec_boundary}, we find the following expressions for
the boundary conditions:
\begin{equation}
  \frac{\rmd\phi(R)}{\rmd r} = -\frac{\delta}{\Omega}
  \left\{ x^2 \left[G(x) - \frac{xg(x)}{1-\phi} \right]
  \left(\frac{\phi}{1-\phi}\right)^2
  + 2\chi\phi + \mu \right\}
\end{equation}
which replaces equation~(\ref{bc1}), and
\begin{eqnarray}
  && \frac{\phi}{1-\phi} = \frac{\chi[(1-\phi)^2+\rho x^2]}{x^3g(x)}
\\
  && \frac{ 2[(1-\phi)G(x)+xg(x)] [(1-\phi)^2 + \rho x^2]^2 } {[x^2g(x)]^2}
  \nonumber\\
  && - \frac{8\eps\delta}{\chi R} \frac{(1-\phi)^2+\rho x^2}{xg(x)}
  = \frac{\mu}{\mu_{\rm bin}}(1-\phi),
\end{eqnarray}
instead of equations (\ref{bc2}) and (\ref{bc3}).

The modified expressions derived in this Appendix were used
to produce Figs.\ \ref{fig_belt}--\ref{fig_profile_diagram}.




\begin{onecolumn}

\begin{figure}[tbh]
\centerline{\resizebox{0.45\textwidth}{!}{\includegraphics{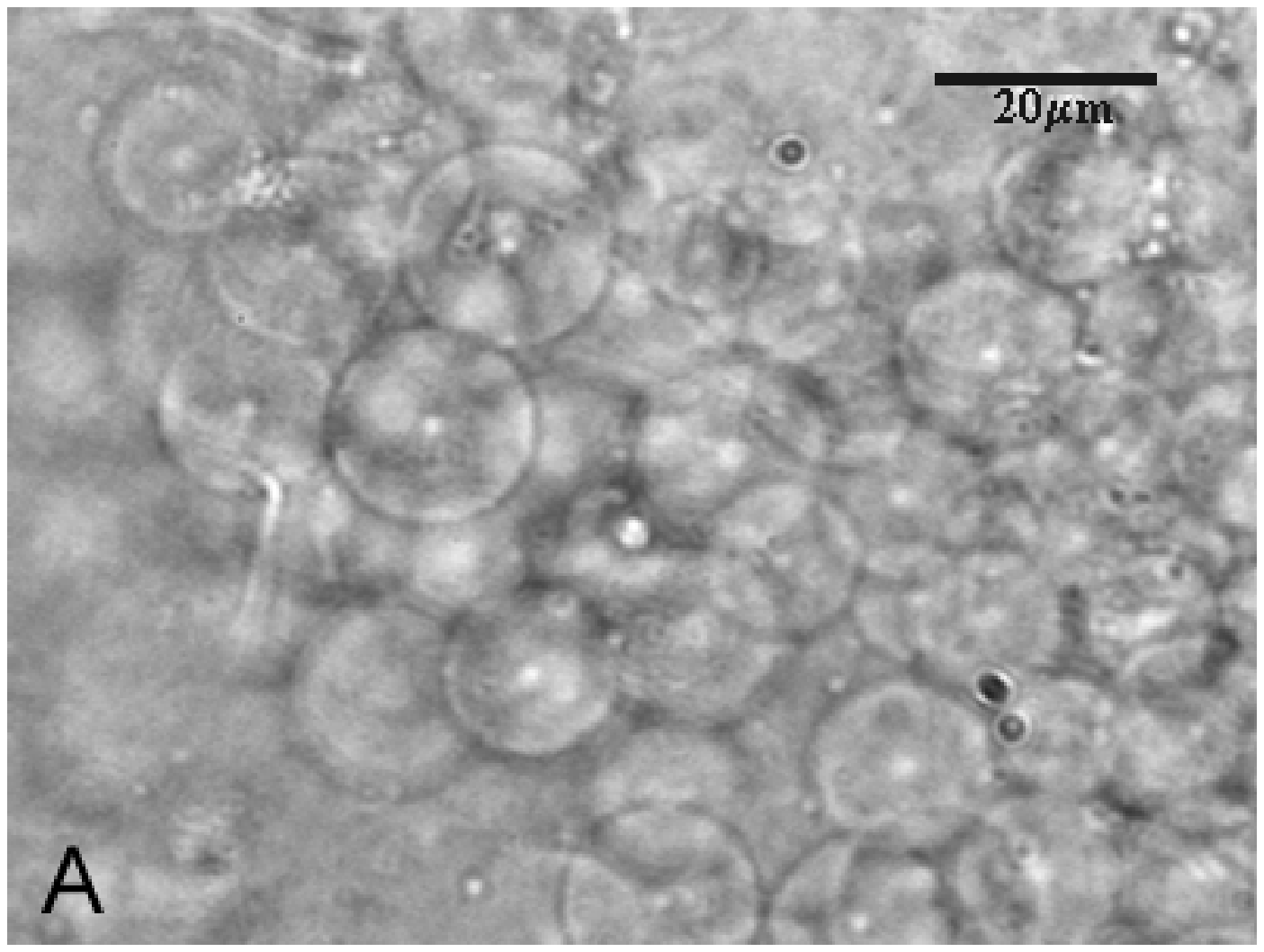}}
\hspace{.5cm}
\resizebox{0.45\textwidth}{!}{\includegraphics{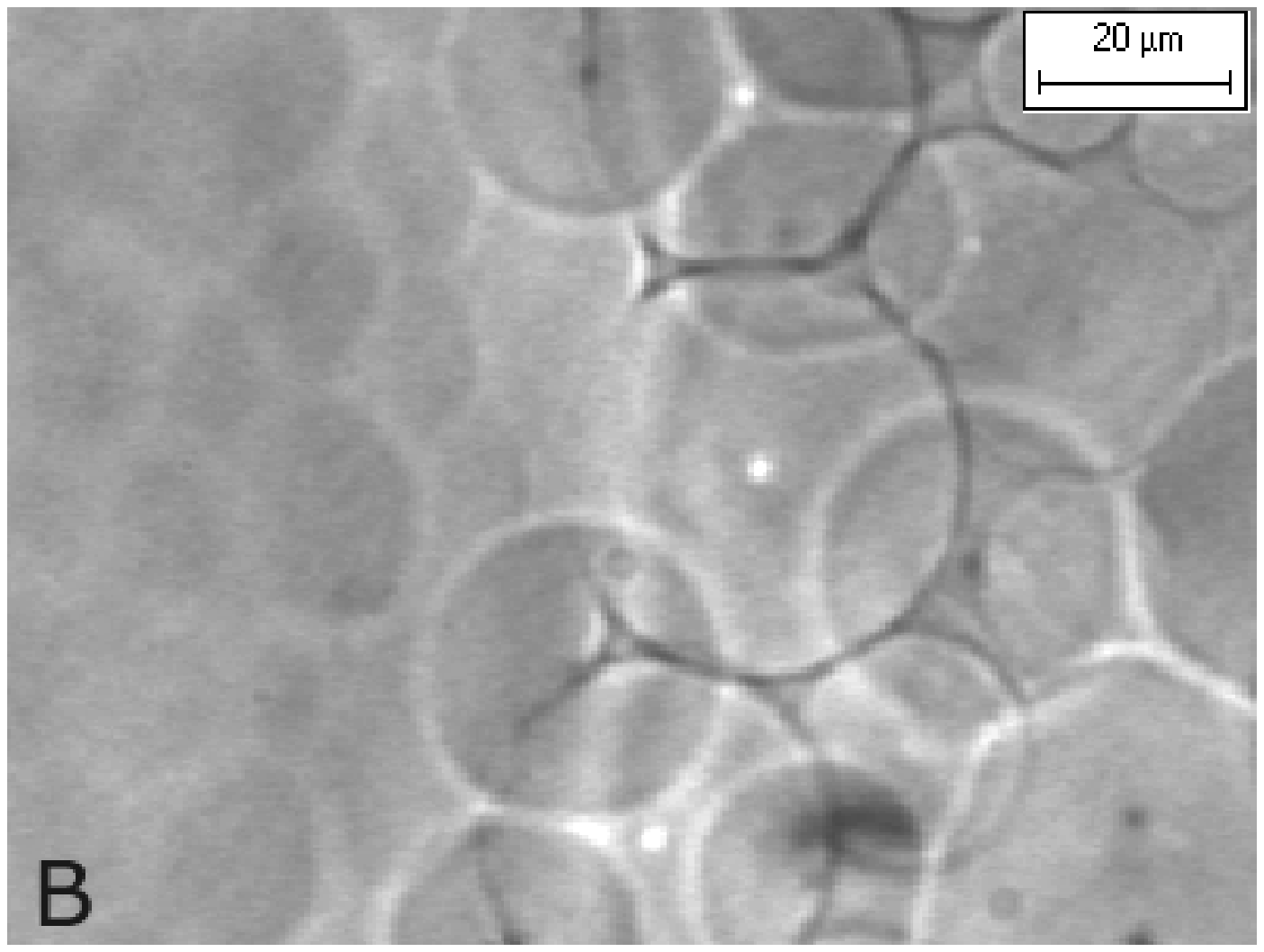}}}%
\caption[]{Formation of `onion cores' upon dilution. 
(A) SDS--octanol--brine system \cite{Buchanan2}. 
(B) AOT--brine system \cite{Buchanan1}. 
(Pictures courtesy of M. Buchanan.)}%
\label{fig_core_exp}
\end{figure}

\begin{figure}[tbh]
\centerline{\resizebox{0.45\textwidth}{!}{\includegraphics{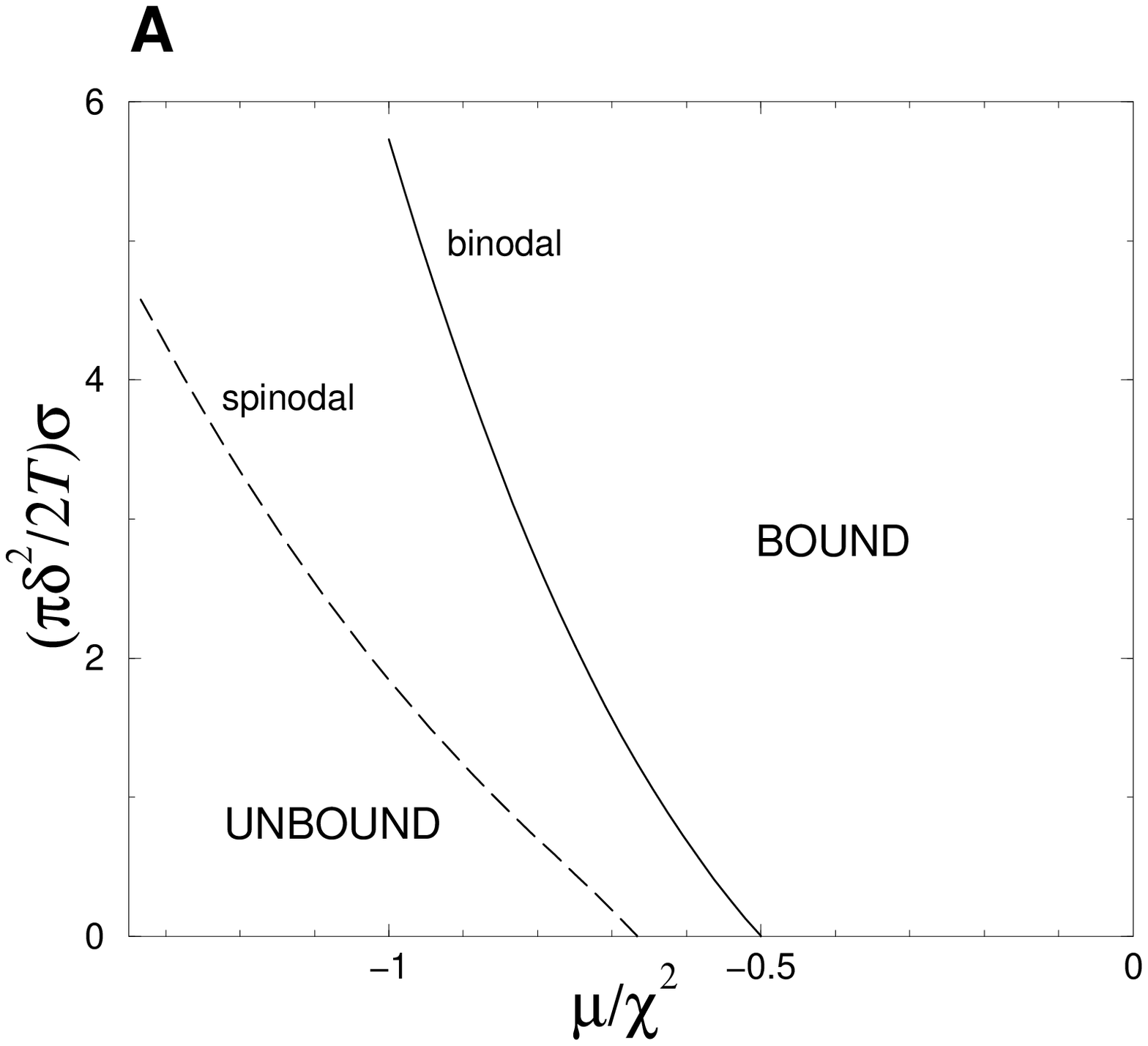}}
\resizebox{0.45\textwidth}{!}{\includegraphics{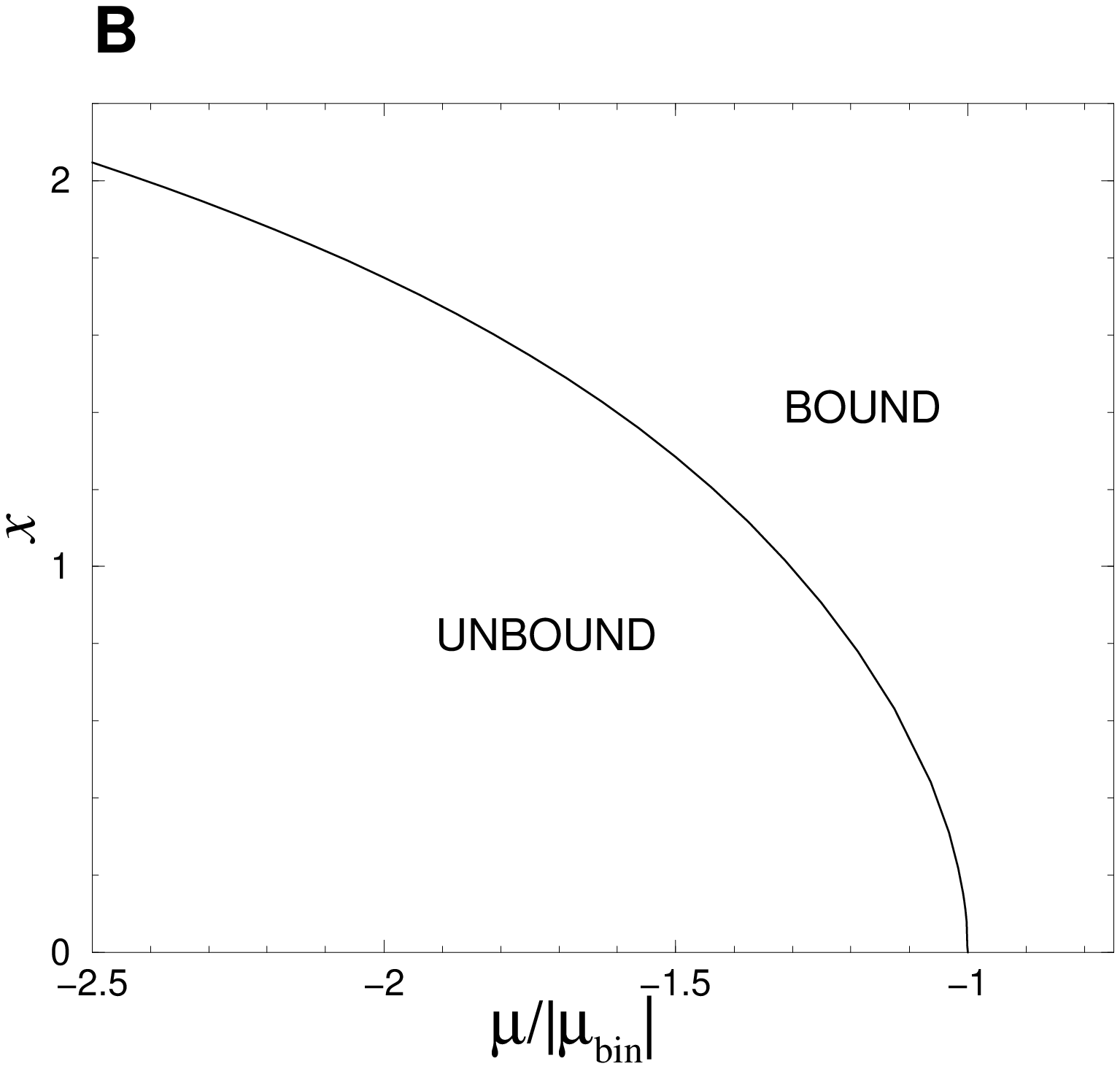}}}%
\caption[]{Phase diagram of a lamellar phase with tension
[Eq.~(\ref{tense_transition})]. (A) Binodal and spinodal
tension--chemical potential lines. (B) Values of $x(\mu)$
corresponding to the binodal.}%
\label{fig_tense_diag}
\end{figure}

\begin{figure}[tbh]
\centerline{\resizebox{0.5\textwidth}{!}{\includegraphics{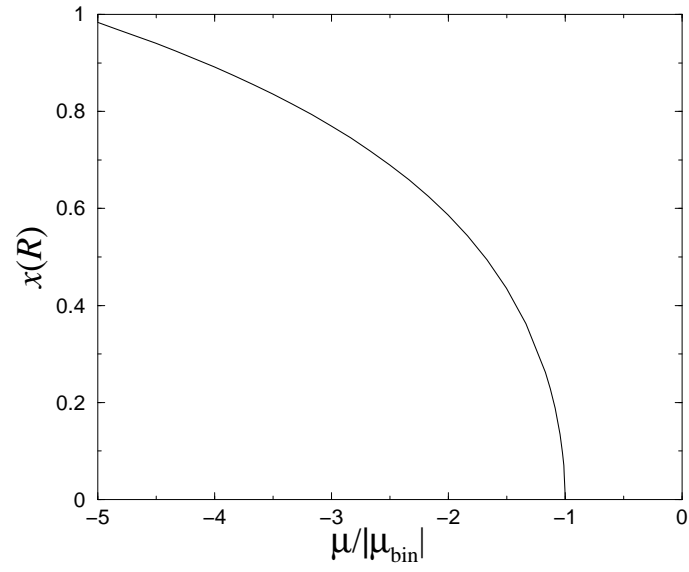}}}%
\caption[]{Boundary value of $x$ as a function of chemical
potential [Eq.~(\ref{bc3}) with $\chi=0.1$, $\eps=0.1$ and
$\delta/R=10^{-3}$].}%
\label{fig_xR}
\end{figure}

\begin{figure}[tbh]
\vspace{1cm}
\centerline{\resizebox{0.5\textwidth}{!}{\includegraphics{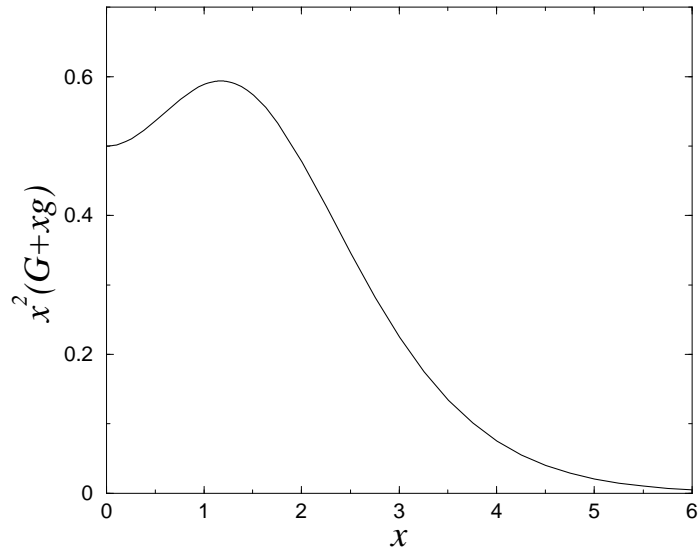}}}%
\caption[]{Dependence of pressure on $x$ for given $\phi$ [cf.\
Eq.~(\ref{pressure})].}%
\label{fig_pressure}
\end{figure}

\begin{figure}[tbh]
\centerline{\resizebox{0.5\textwidth}{!}{\includegraphics{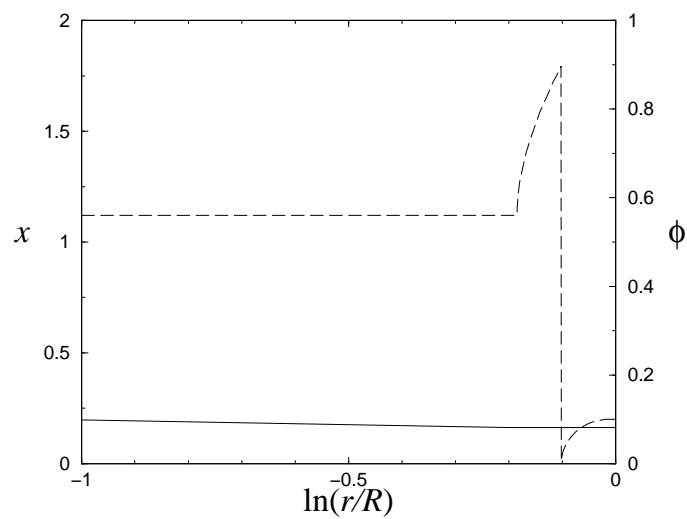}}}%
\caption[]{`Belt' profile exhibiting a narrow region of high
tension. Solid line --- $\phi(r)$, dashed
--- $x(r)$. Parameters: $\chi=0.1$, $\eps=0.1$,
$\delta/R=10^{-3}$, $\Omega/R^2=0.01$, $\mu/|\mu_{\rm
bin}|=-1.1$.}%
\label{fig_belt}
\end{figure}

\begin{figure}[tbh]
\centerline{
\resizebox{0.55\textwidth}{!}{\includegraphics{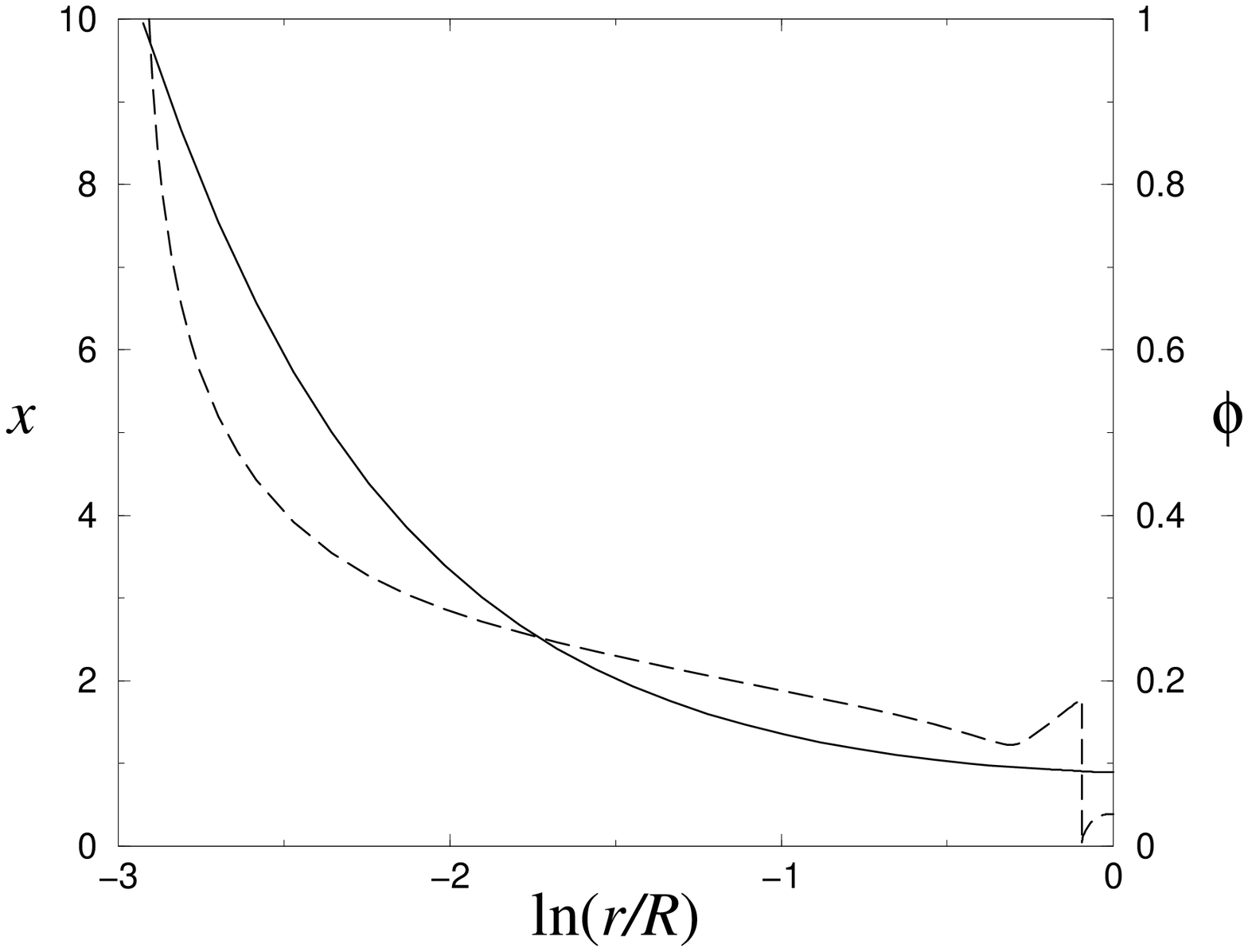}}%
\hspace{.2cm}
\resizebox{0.4\textwidth}{!}{\includegraphics{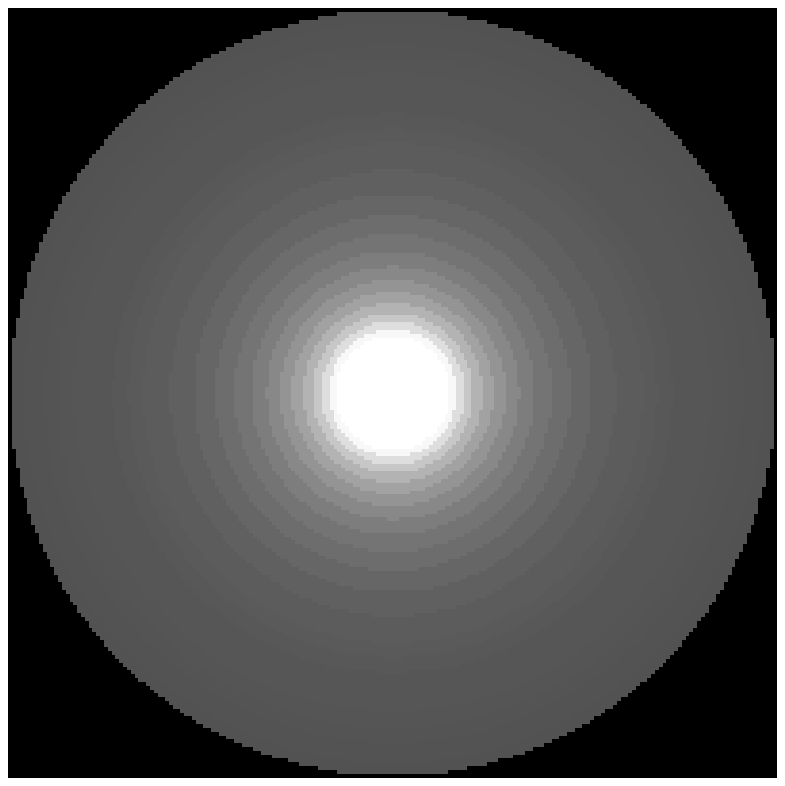}}}%
\caption{`Core' profile. Left --- volume fraction (solid) and
tension (dashed) profiles; note the logarithmic scale of position.
Right --- the volume fraction profile redrawn as a density plot.
Parameters as in Fig.~\ref{fig_belt} except $\mu/|\mu_{\rm
bin}|=-1.4$}%
\label{fig_core}
\end{figure}

\begin{figure}[tbh]
\centerline{\resizebox{0.5\textwidth}{!}{\includegraphics{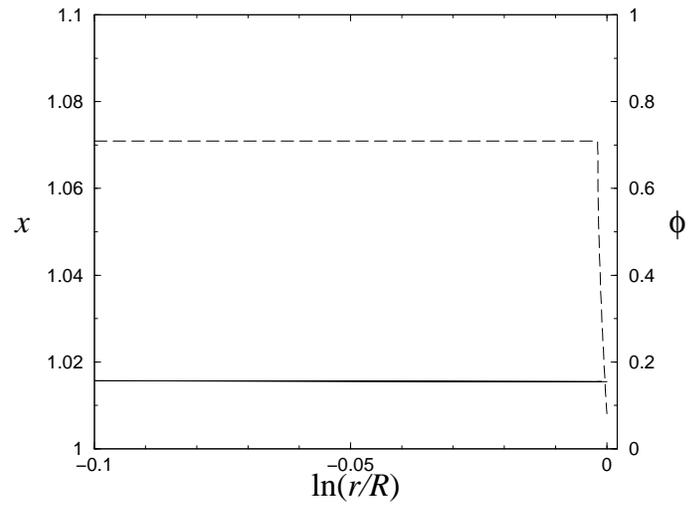}}}%
\caption{`Uniform' profile exhibiting a continuous tension
profile. Parameters as in Fig.~\ref{fig_belt} except
$\mu/|\mu_{\rm bin}|=-6$.}%
\label{fig_uniform}
\end{figure}

\begin{figure}[tbh]
\centerline{
\resizebox{0.45\textwidth}{!}{\includegraphics{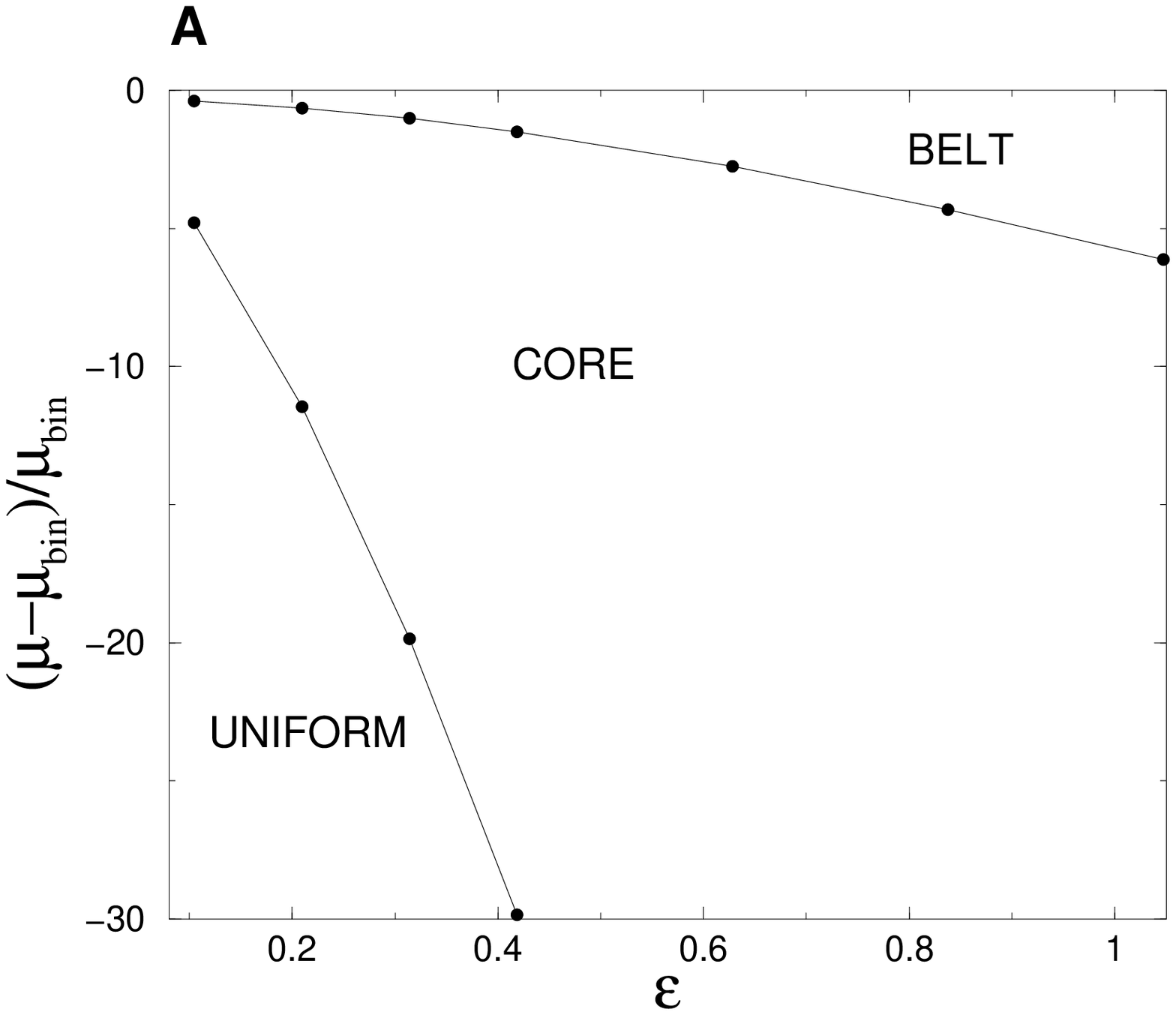}}
\resizebox{0.45\textwidth}{!}{\includegraphics{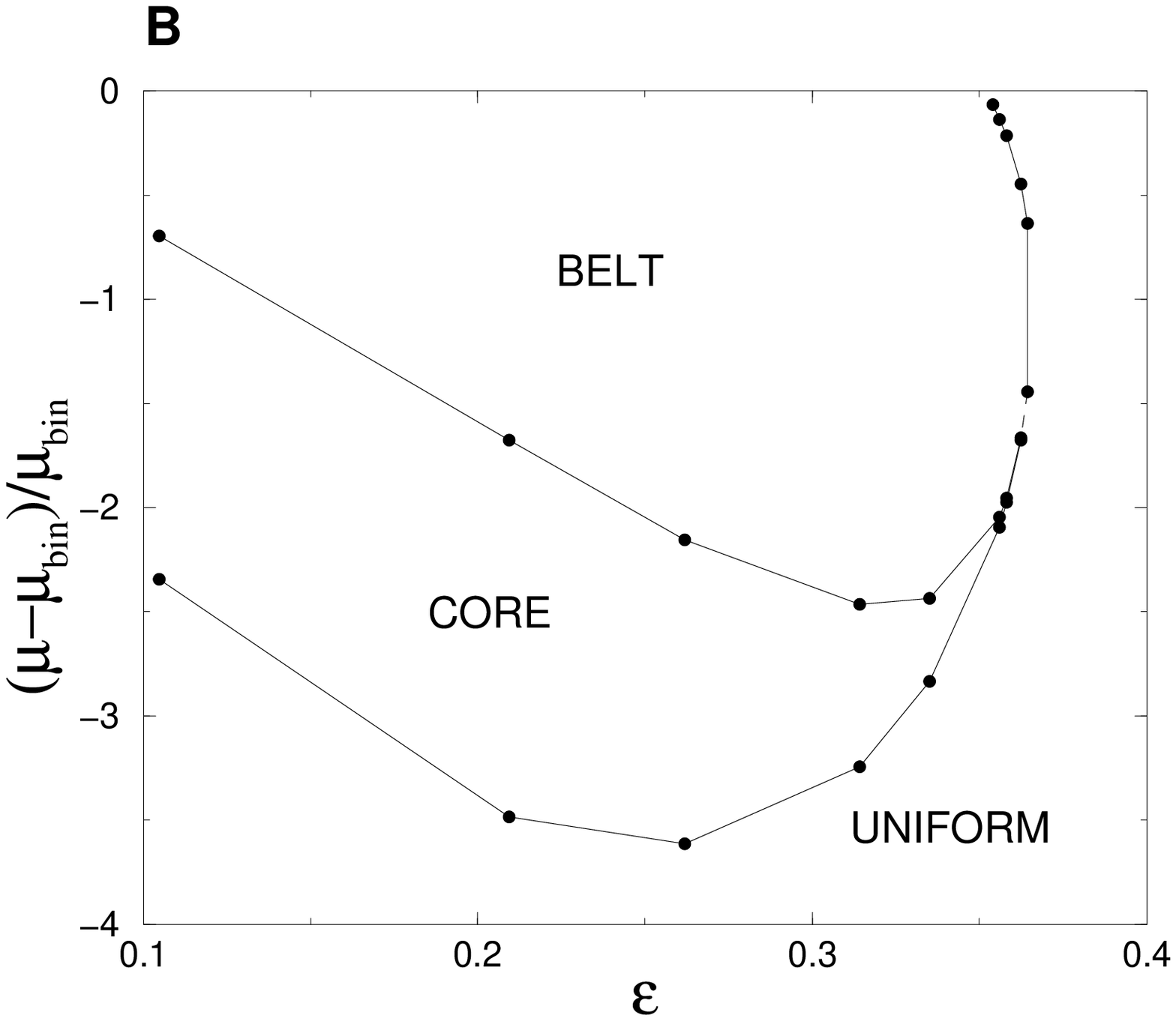}}}
\centerline{
\resizebox{0.45\textwidth}{!}{\includegraphics{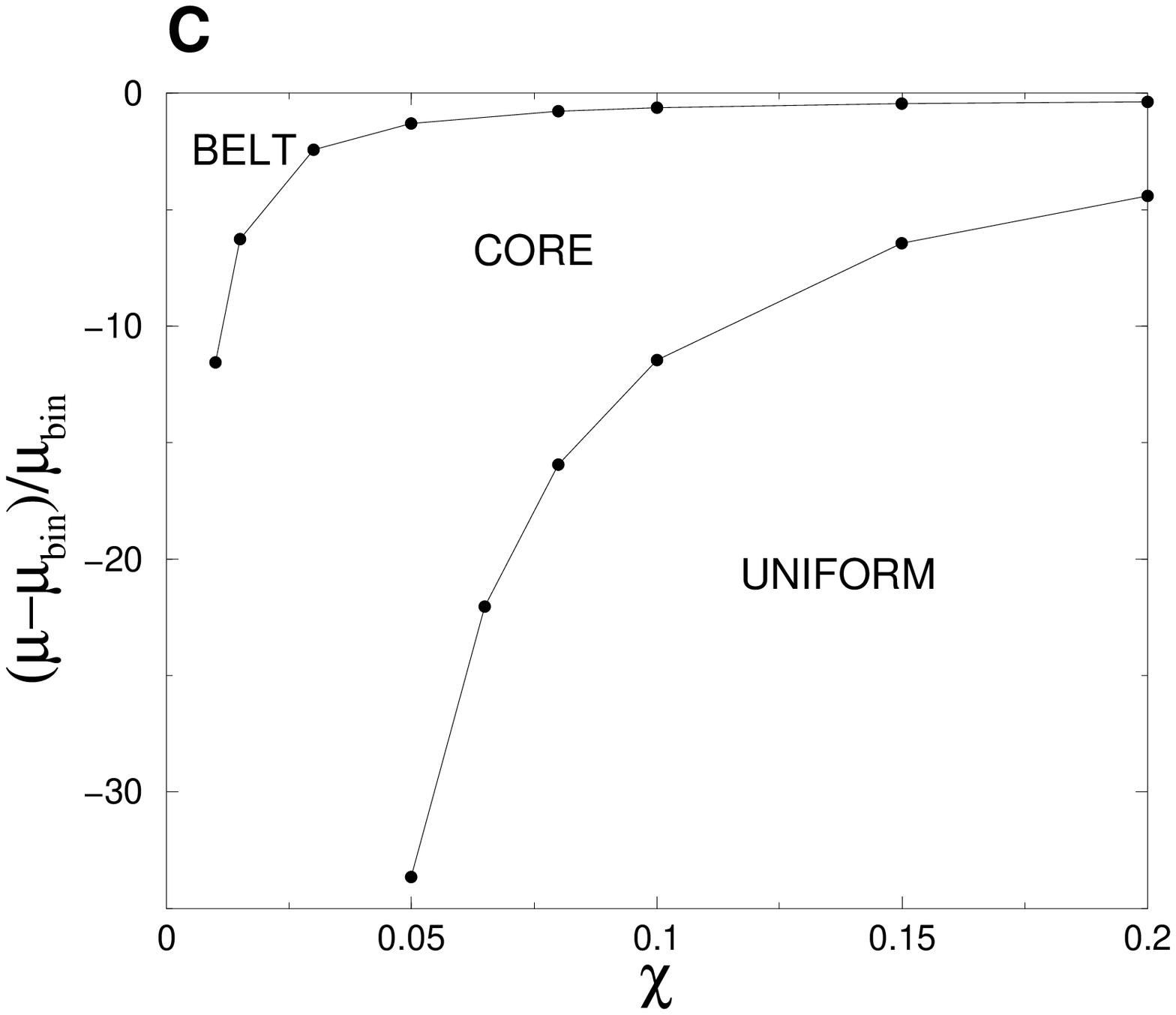}}}%
\caption[]{`Phase diagrams' of profile types encountered during
onion swelling. (a) Dependence on $\eps\sim\kappa/T$. Values of
parameters are $\chi=0.1$, $\delta/R=10^{-3}$, and
$\Omega/R^2=0.01$. (b) Same as (a) except $\Omega/R^2=0.1$. (c)
Dependence on $\chi$. Values of parameters are $\eps=0.21$,
$\delta/R=10^{-3}$, and $\Omega/R^2=0.01$.}%
\label{fig_profile_diagram}
\end{figure}


\end{onecolumn}



\begin{thebibliography}{}

\bibitem{Benshaul}
\textit{Micelles, Membranes, Microemulsions, and Monolayers},
edited by W. M. Gelbart, A. Ben-Shaul and D. Roux
(Springer-Verlag, New York, 1994).

\bibitem{GompperSchick}
G. Gompper and M. Schick, \textit{Self-Assembling Amphiphilic
Systems}, in \textit{Phase Transitions \& Critical Phenomena},
edited by C. Domb and J. L. Lebowitz (Academic Press, London,
1994).

\bibitem{Roux1}
O. Diat and D. Roux, J. Phys. II France \textbf{3}, 9 (1993).

\bibitem{Roux2}
O. Diat, D. Roux and F. Nallet, J. Phys. II France \textbf{3},
1427 (1993).

\bibitem{Roux3}
D. Roux, F. Nallet and O. Diat, Europhys. Lett. \textbf{24}, 53
(1993).

\bibitem{Roux4}
O. Diat, D. Roux and F. Nallet, Phys. Rev. E \textbf{51}, 3296
(1995).

\bibitem{Roux5}
For a recent review, see D. Roux, in \textit{Soft and Fragile
Matter, Nonequilibrium Dynamics, Metastability and Flow}, edited
by M. E. Cates and M. R. Evans (IOP Publishing, Bristol, 2000).

\bibitem{encapsulation}
J. Arrault, C. Grand, W. C. K. Poon and M. E. Cates, Europhys.
Lett. \textbf{38}, 625 (1997).

\bibitem{ZilmanGranek}
A mechanism involving undulation instability under shear has been
recently suggested. A. G. Zilman and R. Granek, Eur. Phys. J. B
\textbf{11}, 593 (1999).

\bibitem{Lekker}
E. van der Linden, W. T. Hogervorst and H. N. W. Lekkerkerker,
Langmuir \textbf{12}, 3127 (1996).

\bibitem{Panizza}
P. Panizza, D. Roux, V. Vuillaume, C.-Y. D. Lu and M. E. Cates,
Langmuir \textbf{12}, 248 (1996).

\bibitem{Buchanan1}
M. Buchanan, J. Arrault and M. E. Cates, Langmuir \textbf{14},
7371 (1998).

\bibitem{Buchanan2}
M. Buchanan, S. U. Egelhaaf and M. E. Cates,
Colloid Surf. A, submitted.
M. Buchanan, Ph.D. thesis,
Univsersity of Edinburgh, 1999.

\bibitem{Safran}
S. A. Safran, \textit{Statistical Thermodynamics of Surfaces,
Interfaces, and Membranes} (Addison-Wesley, New York, 1994),
Chapter 8.

\bibitem{inextensible}
Ref.~\cite{Safran}, Chapter 6.

\bibitem{passages}
W. Harbich, R.-M. Servuss and W. Helfrich, Z. Naturforsch. A
\textbf{33}, 1013 (1978). X. Michalet, D. Bensimon and B.
Fourcade, Phys. Rev. Lett. \textbf{72}, 168 (1994). T. Charitat
and B. Fourcade, J. Phys. II France \textbf{7}, 15 (1997).

\bibitem{Kittel}
See, \eg C. Kittel and H. Kroemer, \textit{Thermal Physics},
(Freeman, New York, 1980), Chapter 10.

\bibitem{Andelman95}
M. Seul and D. Andelman, Science \textbf{267}, 476 (1995).

\bibitem{Milner92}
S. T. Milner and D. Roux, J. Phys. I. France \textbf{2}, 1741
(1992).

\bibitem{Helfrich1}
W. Helfrich, Z. Naturforsch. \textbf{33a}, 305 (1977). W. Helfrich
and R.-M. Servuss, Nuovo Cimento \textbf{3D}, 137 (1984).

\bibitem{Sornette}
D. Sornette and N. Ostrowsky, in Ref.~\cite{Benshaul}.

\bibitem{Lipowsky89}
R. Lipowsky and B. Zielinska, Phys. Rev. Lett. \textbf{62}, 1572
(1989).

\bibitem{Nelson_Leibler}
S. Leibler, in \textit{Statistical Mechanics of Membranes and
Surfaces}, edited by D. Nelson, T. Piran and S. Weinberg (World
Scientific, Singapore, 1989).

\bibitem{LipowskyLeibler}
R. Lipowsky and S. Leibler, Phys. Rev. Lett. \textbf{56}, 2541
(1986). S. Leibler and R. Lipowsky, Phys. Rev. B \textbf{35}, 7004
(1987).

\bibitem{RG_tension}
R. Lipowsky, in \textit{Structure and Dynamics of Membranes},
edited by R. Lipowsky and E. Sackman, \textit{Handbook of
Biological Physics} (Elsevier, Amsterdam, 1995).

\bibitem{Netz95}
R. Netz and R. Lipowsky, Europhys. Lett. \textbf{29}, 345 (1995).

\bibitem{Seifert}
U. Seifert, Phys. Rev. Lett. \textbf{74}, 5060 (1995).

\bibitem{Langer}
See, \eg J. S. Langer in \textit{Solids Far From Equilibrium},
edited by C. Godr\`{e}che (Cambridge University Press, Cambridge,
1991).



\end{thebibliography}
\end{document}